\pdfoutput=1
\documentclass[11pt]{article}

\usepackage[]{naacl2021}
\usepackage{multibib}

\usepackage{times}
\usepackage[linesnumbered,ruled,vlined]{algorithm2e}

\usepackage{latexsym}
\usepackage{graphicx}
\usepackage[compact]{titlesec}
\usepackage[T1]{fontenc}
\usepackage[utf8]{inputenc}

\usepackage{microtype}

\title{A Comprehensive Attempt to Research Statement Generation}

 \author{Wenhao Wu \and Sujian Li \\
        Key Laboratory of Computational Linguistics, MOE, Peking University \\
        \texttt{\{waynewu,lisujian\}@pku.edu.cn}}
\begin{document}
\maketitle
\begin{abstract}
For a researcher, writing a good research statement is crucial but costs a lot of time and effort. To help researchers, in this paper, we propose the research statement generation (RSG) task which aims to summarize one's research achievements and help prepare a formal research statement. For this task, we conduct a comprehensive attempt including corpus construction, method design, and performance evaluation.
First, we construct an RSG dataset with 62 research statements and the corresponding 1,203 publications. Due to the limitation of our resources, we propose a practical RSG method which identifies a researcher's research directions by topic modeling and clustering techniques and extracts salient sentences by a neural text summarizer.
Finally, experiments show that our method outperforms all the baselines with better content coverage and coherence\footnote{We release our RSG dataset and benchmark method at an anonymity link.}.
\end{abstract}
\begin{figure*}[t]
\centering
\includegraphics[scale=0.4]{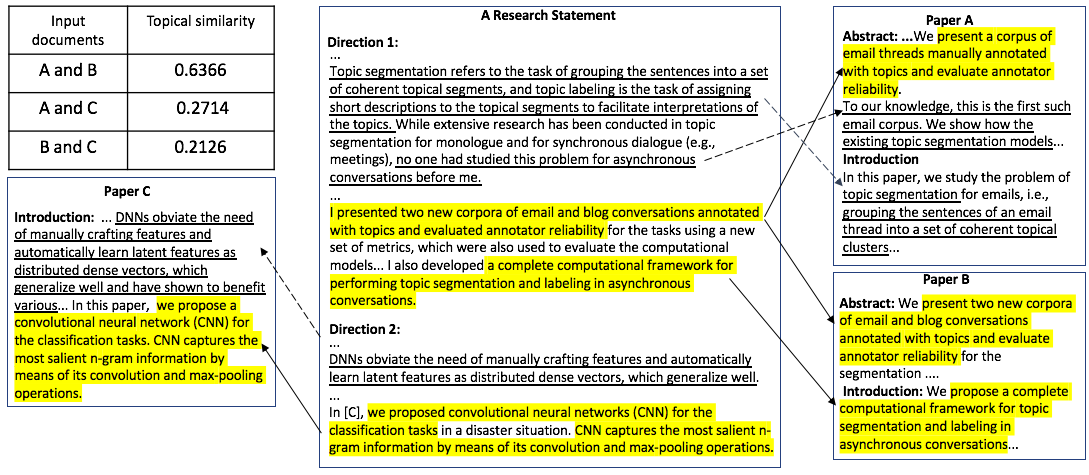}
\caption{Illustration of a part of a research statement which involves two directions and the corresponding three papers. \textbf{Paper A }and \textbf{B} belong to\textbf{ Direction 1} and \textbf{Paper C} belongs to \textbf{Direction 2}. The highlighted text and directed edges intuitively show the relations between the directions and the papers. %Highlighted sentences matched with a solid arrow are directly copied matches, and underlined sentences matched with dash line are similar in semantic.
}
\label{fig:illus}
\end{figure*}
\section{Introduction}
A research statement is a summary of research achievements and a proposal for future research.  
A  research statement can assist in a job application and position promotion for an applicant.
However, writing a research statement is tricky and requires much effort and time to collect evidence and summarize achievements from a lot of academic papers.
Thus, we propose the task of automatic research statement generation (RSG), which aims to aid a researcher in preparing a formal research statement. 
In this paper, we mainly focus on automatically summarizing one's achievements, which constitutes the main content of a research statement, according to his/her publications.

Here, the RSG task takes multiple academic papers as the input and a brief overview of one's achievements as the output, which is similar to the multi-document summarization (MDS) task.
Different from the general-purpose MDS,
which is usually oriented with one event and conducted on the news datasets ~\cite{yasunaga2017graph,gupta2012multi}, the RSG task focuses on multiple research directions and summarizes their content.
Figure \ref{fig:illus} shows a part of a real research statement and three papers. From this figure, we can see that \textit{Paper A} and \textit{Paper B} belong to the direction of \textit{topic segmentation} and \textit{Paper C} belongs to the direction of \textit{deep learning}.
The corresponding research statement describes the work in the two directions which have involved the three papers. 
%Compared to the general MDS task, 
At the same time, 
%one challenging problem of RSG is that 
it is difficult for the RSG task to collect a sizeable corpus  for model supervision as only a few senior researchers would like to release their research statements.
Thus, in such a low-resource scenario of RSG, %to automatically generate a research statement, 
two key problems are faced with:
(1) How to  well represent papers and group them into different research directions; 
%(1) How to select salient sentences in a low resource situation.
(2) How to get some salient sentences  and organize them with better coherence  for describing each research direction.
%when generating the research statement.
%(2) How to select a large number of sentences with low redundancy.

According to the analysis above, in this paper, 
we propose a practical RSG method which exploits unsupervised techniques  to determine a researcher's main research directions and then summarizes his/her achievements for each direction with the help of external resources.
Specifically, to well represent 
%the topic distribution of 
the papers, we adopt the unsupervised neural topic model to explore the latent topic representations of text. Based on the topic representation, we cluster the papers into several directions using the affinity propagation (AP)  clustering method, %and select important sentences for each research direction.
%Here, we adopt the affinity propagation (AP)  clustering method %based on the topic representation, 
because this method does not require to predefine the number of clusters.
To identify one's achievements, we adopt the BERT-based method B\textsc{ertsum} trained on an external scientific summarization corpus to pick out the sentences which are essential to expressing one's achievements in all directions. %from the corresponding cluster of papers. 
Finally, we reorder the selected sentences based on their topic representations to promote the coherence of the generated statement.
 
We also crawled and compiled a dataset of 62 statements with the corresponding 1,203 publications for the RSG task. 
With this corpus, we evaluate our proposed benchmark method.
Our main contributions are summarized as follows: 
\begin{itemize}
\item We first propose the research statement generation task which can help an applicant prepare his/her formal research statement.
\item We design a benchmark method which can automatically determine one's main research directions and summarize one's achievements for each direction.
\item We build a small dataset of RSG which is helpful to the research in the field.
\end{itemize}
\section{RSG Task Definition and Dataset}
\subsection{Task Definition}
In this work, we define the RSG task as generating a research statement according to a set of papers.
Formally,
for a researcher $A$ with a paper set containing  $n_A$ papers  $P_A=\{p_1,p_2,p_3, \dots p_{n_A}\}$, 
the RSG task aims to output the statement $R_A$. Then, RSG can be formalized as the function
$R_A=f(P_A)$.
%generating a research statement  $R_A$, which summarizes all the papers in $P_A$ published by $A$. 
%Here, we formalize the RSG task as a function $R_A=f(P_A)$.
To the best of our knowledge, we are the first to formally propose the RSG task.
%Thus, to implement and evaluate this task, the first key thing  is to construct an RSG dataset. 
\subsection{Dataset Construction}
As the first attempt at RSG, we collect and compile a dataset mainly for evaluation.
It is difficult to construct such a corpus because only a small number of researchers post their research statements online, and their published papers scatter in different conferences and journals some of which may be inaccessible. %which is hard for automatic collection.
 We have collected 110 research statements written by different researchers using  search engines. 
In these statements,  some of them are not suitable for our task; for example, the master students' statements for Ph.D. applications only contain a small number of publications and mainly focus on future plans.
After we remove all the informal or unqualified statements manually,  only 62  research statements are kept and belong to different research fields of computer science, including Natural Language Processing (NLP), Information Security, etc.
We use a Java library Cermine \cite{tkaczyk2015cermine}  to convert the PDF files into the XML format for getting high-quality texts.
Next, based on the  content of each statement, we use a crawler to automatically search for the corresponding papers which appear in the reference list of each statement and download them.
%The file conversion noises and the scattering publications make it hard to  crawl all the papers. %and thus we can not assure the completeness of the paper lists.
%However, the incompleteness of publications doesn't influence the overall performance much because one researcher's most contributed publications can usually be successfully downloaded by our crawler. % and we further verify this in section \ref{ARS}.

Our final RSG dataset contains 62 research statements written by 62 distinct researchers with their corresponding 1,203 publications.
On average, each statement is composed of  89.4 sentences or 1,967 words.
To add an explanatory note, the scale of our corpus is not large, but is comparable to that of some  multi-document summarization (MDS) datasets such as DUC or TAC MDS corpus each of which is not up to 50 topics with the average of 25 documents per topic~\cite{hoa2006overview,dang2008overview,dang2009overview}, whose small scale is mainly caused by costly manual summarization of multiple documents.
In addition, compared to previous MDS task with much shorter reference summary (a maximum of 250 words),
%with much smaller single topic inputs and shorter references,
RSG requires to output a longer overview and is more challenging.
\section{Analysis of Research Statements}
We analyze the content coverage of each research statement over its corresponding paper set. 
Through this analysis, we explore the upper bound of extractive methods and provide more inspiration for further research.

As we know, the content of a research statement is not necessarily covered by the corresponding research papers.
%In many statements, researchers introduce their contributions along with others' evidence which cites his/her work.
In many statements, researchers may summarize their contributions using new words which are different from those in the papers.
Noticing the possible lack of information that exists in the research papers, we are curious about the effectiveness of using the extractive method.  %to generate a statement .
%At the same time, it is also important to see which sections in the papers carry more information which can be summarized into the statements.
%Thus, we study the following two questions: (1) whether it is enough only to use a researcher's papers as source documents. (2) whether it is appropriate to  use some sections for extraction.
\begin{table}
\label{table:analyse}
\footnotesize
\centering
\begin{tabular}{ccccc}

    \hline
    Section&R-L&R-2&R-1&Sentence Ratio \\
     \hline
     \textit{fullPaper}\footnotemark[1]&$\approx 0.98$&0.9984&0.9999&75.41\\
     \textit{Abs+Intro}& 0.7491&0.2950&0.7630&9.38\\
     \textit{Abs}&0.4757&0.1280&0.4878&1.48\\
    \hline
\end{tabular}

\caption{Information coverage of different sections over research statements}
\vspace{-1.5em}
\end{table}

Here we concatenate certain parts of papers as a forged research statement similar to \cite{DBLP:journals/corr/VermaL17}
and use ROUGE Recall to evaluate information coverage over the ground-truth statement.
We also define the metric of \textit{Sentence Ratio} which calculates the ratio between
the sentence numbers of the forged statement and the ground truth. 
High ROUGE scores with a low \textit{Sentence Ratio} is the ideal result. 
Usually, we seek a balance between the ROUGE and \textit{Sentence Ratio}.
From the results in  Table 5, we can see that the combination of full papers (\textit{fullpaper}) nearly covers all the unigrams and bi-grams of the statements but the \textit{Sentence Ratio} is very high.
Meanwhile, All the abstract plus introduction sections (\textit{Abs+Intro}) contain about 80\% of the unigrams on average, and the \textit{Sentence Ratio} is about 1:9.38.
Compared to \textbf{Abs+Intro}, it is more difficult to use full paper (with Sentence Ratio of 1:75) as the  source text  for sentence selection.
Combining all the abstracts (\textit{Abs}) only covers less than a half of the unigrams in the statement and cannot provide enough information for further summarization.
%gives a similar number of sentences compared to the target statement, but only contains half of the unigrams in the statement, which is not enough for further summarization.
%Combining all the abstracts (Abs) gives a similar number of sentences compared to the target statement, but only contains half of the unigrams in the statement, which is not enough for further summarization.
Thus, \textit{Abs+Intro} can be seen as a balance to serve as the source text for RSG.
It is also noted that \textit{Abs+Intro} performs poorly with respect to ROUGE-2, meaning that many ground-truth sentences do not directly come from these two parts. This conforms to our statistics: 15.2\%  of ground-truth sentences come from the abstracts, 25.3\% from the Introduction sections and 59.5\% from other sections.
This means only using the content of Abstract and Introduction as source text also limits the upper bound of our extractive method.
How to make full use of other sections %with fewer efforts 
will a  future  focus.
\section{Our RSG Method}
\begin{figure}
    \centering
    \includegraphics[scale=0.16]{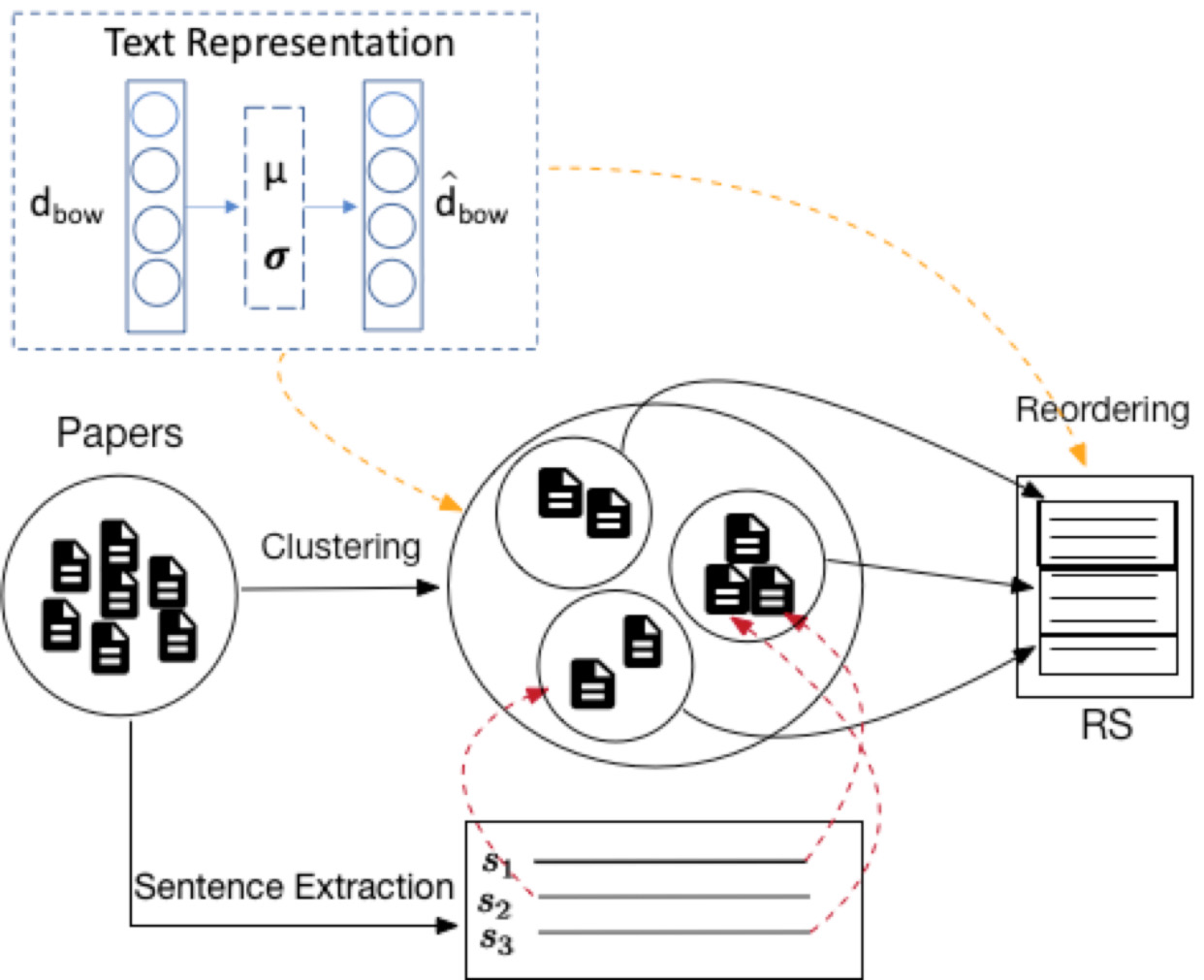}
    \caption{Overall Architecture of Our RSG Method.}
%    \caption{Our pipeline for research statement generation.}
    \label{fig:pipeline}
    \vspace{-1em}
\end{figure}

The overall architecture of our RSG method is shown in Figure \ref{fig:pipeline}.
First, we adopt the neural topic model (NTM) to represent text based on which we cluster papers into different research directions, and select and order the sentences which can summarize the achievements.

%\subsection{Document Clustering}
\subsection{Topic Representation of Text}
A research statement is concisely organized by research directions
each of which is composed of papers involving similar fine-grained topics. 
%with each direction summarizing papers with similar fine-grained topics.
For example,  papers in the direction of \textit{text generation} may involve the topics like \textit{seq2seq methods} or  \textit{abstractive summarization}.
With such idea, we explore topic modeling methods and 
adopt the variational autoencoder based neural topic model (NTM)~\cite{miao2017discovering} to discover latent topics which are derived from word co-occurrence.
Compared to previous Bayesian topic models such as  LDA \cite{blei2003latent}, NTM
does not rely on
much expertise involvement such as predefining many prior hyper-parameters, but provides 
parameterisable distributions which permit training by backpropagation.

We predefine  $K$  fine-grained latent topics and represent each document as a distribution over the $K$ topics by NTM.

%Illustrated in Neural topic model part
As in Figure \ref{fig:pipeline}, for each document with a bag of words $d_{bow}$ as input, the neural topic model uses the neural perceptrons to separately encode it into two prior parameters $\mu$ and $\sigma$: %which will be used to induce the intermediate topic representation $z$:
\begin{equation}
\mu  = ReLu(W_1^Td_{bow}), \sigma = ReLu(W_2^Td_{bow})
\end{equation}
where $W_1$ and $W_2 \in R^{V \times H}$ represent the parameters of two different neural perceptrons, $V$ is the vocabulary size % of $d_{bow}$ 
and $H$ is the dimension of the prior parameters. %size of the prior parameters.
Similar to LDA, we assume each topic is represented as a distribution $\phi$ over words and each document in the corpus is represented as a $K$-dim topic mixture distribution $\theta$.
Following \cite{miao2017discovering} we construct $\theta$ by Gaussian softmax:
\begin{equation}
z \sim N(\mu,\sigma^2), 
\theta = softmax(W_t^Tz)
\end{equation}
where $W_t \in R^{H\times K}$ is a linear transformation and $z$ is sampled from normal distribution according to 
$\mu$ and $\sigma^2$.
The decoding process reconstructs the input as $\hat{d}_{bow}$ by projecting the topic mixture distribution $\theta$ through topic-word distribution $\phi$ :
\begin{equation}
\hat{d}_{bow}=  \phi^{T} \theta 
\end{equation}
Here, $\phi \in R^{K \times V}$ is a learnable parameter. For training, we apply VAE's loss function using reconstruction result and prior parameters according to \cite{miao2017discovering}. Finally, with NTM we can get a topic representation for each document.
\subsection{Research Direction Determination by Clustering Papers}
To automatically determine  which research directions a researcher have made achievements in, we choose to cluster the papers based on their topic representations derived from NTM,
with the assumption that documents in the same direction should share similar distributions over latent topics.

%collect a researcher's achievements on each research direction and clustering algorithm to 
The number of research directions varies from researcher to researcher, and  is usually proportional to the number of the published papers.
Here we apply Affinity Propagation (AP) \cite{frey2007clustering}, a non-parametric clustering algorithm,  to cluster a researcher's papers. 
Unlike the widely used $K$-means clustering method, AP does not require a predefined number of clusters and thus the number of research directions can be determined automatically. 
%Thus, we adopt the Affinity Propagation (AP) method~\cite{frey2007clustering}, which does not require a predefined cluster number, to cluster a researcher's papers.
%researchers cluster documents using methods like $K$-means, which requires a predefined cluster number.

%We then cluster the paper set $P$ into several directions by their topical similarities using Affinity Propagation (AP) \cite{frey2007clustering}.

%In our task,  the number of directions varies from researcher to researcher. 

%the number of research directions and statement's length is proportional to the number of papers.
%Thus we apply Affinity Propagation (AP) \cite{frey2007clustering}, a non-parametric clustering algorithm, which does not require predefined number of clusters. 

The AP approach is based on similarities between data points.
We measure the similarity of two different documents by their topical distributions. 
Formally, for two documents  $d_1$ and $d_2$, with their corresponding topic representations $\theta_1$ and $\theta_2$, we define their topical similarity as :
\begin{equation}\label{eq:sim}
T_{sim}(d_1,d_2) = \theta_1^T\theta_2
\end{equation}
In this way, papers with similar topic distribution tend to belong to a same research direction.
%Illustrated in the upper left table of Figure \ref{fig:illus}, papers similar in direction (A and B) achieve much higher topical similarity. 
\subsection{Sentence Extraction and Statement Generation}
%We select important sentences from different directions in sentence scoring and selection two steps.
Here, we  adopt the state-of-the-art summarization method to extract important sentences and reorder them according to their research directions to compose of the final research statement.
Specifically, we use the state-of-the-art extractive summarization method B\textsc{ertsum} \cite{liu2019fine} which applies BERT \cite{devlin2018bert}
to encode each sentence and calculate its salience score.
%as summarizer.
%First, we use neural text summarization model for modeling sentence importance based on which we extract important sentences.
%Neural network-based summarization models are data-hungry and need thousands of document pairs for training.
Since such neural network-based summarization models are data-hungry and need thousands of documents and their summaries for training, direct training with our RSG dataset suffers from severe data sparsity problem.
%Directly training on our RSG dataset suffer from severe data sparsity problem.
Thus, 
%we adopt the single document summarization technique in sentence extraction.
it is important to obtain a large summarization dataset, which is composed of scientific publications, for training. Fortunately, \citet{collins2017supervised} have recently released a  dataset CSPubSum for scientific summarization, which is created by exploiting the existing resource ScienceDirect\footnote{www.sciencedirect.com}.
Then, we use CSPubSum to train B\textsc{ertsum}  with the cross-entropy loss. 

To extract important sentences which can reflect a researcher's main contributions, we regard the set of published papers as one whole document and segment it into sentences.
Then, we apply the trained B\textsc{ertsum} model to rank the salience of each sentence. With the salience scores, we apply 
the Maximum Marginal Relevance (MMR)  method \cite{carbonell1998use} to reduce redundancy while maintaining to include the most salient sentences into the statement.
%select a sentence at each step based on:
%\begin{equation}
% \argmax_{s_i\notin S}[\lambda \hat{s_i}-(1-\lambda)\max _{s_j\in S}(R_L(s_i,s_j))]
%\end{equation}
%where $R_L$ is ROUGE-L $F_1$ defined as the metric of measuring redundancy, $S$ is current extracted sentence set and $\lambda$ is the balancing factor.

%In reality, when writing a statement, a researcher usually spends more paragraphs on more primary research topics. Length constraints for different directions are trivial and secondary for our task, we left it for future research.
%In our experiments, we extract sentences without direction level length constraints as default.
%, we also experiments extract sentences with average length in every direction for fair comparison with models with different clustering settings.
\SetKwProg{Def}{def}{:}{}
\begin{algorithm}
\scriptsize
 \KwData{Input Cluster-Paper set $C =(P_{1},P_{2},$ \dots$)$\;
  Paper-sentence set $P_{i} = (s_{i,1},s_{i,2}$\dots$)$ contains selected sentences}
  \KwResult{Ordered sentence set $R$ }
  \Def{Coherency($P_a$,$P_b$)}{
  
     return $T_{sim}(P_a[-1],P_b[1])$\;
     }
  Sort sentences in every $P_{i}$ by their orders in original document\;
  Randomly select $m_{th}$ set $P_{m}$ in $C$ \;
  Add all sentences in $P_{m}$ to $R$\;
  Del $P_{m}$ in $C$\;
  \While{$C$ not empty}{
        Find $P_K \in C$  maximize Coherence($P_{m}$,$P_k$)\;
        Add sentences in $P_K$ to $R$\;
        $P_{m}=P_K$\;
        Del $P_k$ from $C$\;
    }
 \caption{Sentence reordering for a cluster}
\end{algorithm}
After sentence selection, we organize the selected sentences into different research directions as illustrated in Algorithm 1.
Here, we restore the sentence to its original paper and utilize the research direction of the paper as the research direction of the sentence.
To reorder the sentences in each research direction, first, 
we gather sentences from the same paper and sort them by their appearing orders in the paper.
%For sentences from different papers, we select the paper whose topics are more uniformly distributed, and first order its sentences which may
Second, we randomly select one paper and sort its sentences  as the first ones in this direction, as we think that a statement can start from summarizing any one paper.
%For the rest sentences,
Next, we order the papers in this direction by greedily choosing the paper which has the maximum topical similarity with the last ordered paper.% in the previous step. 
If one paper is ordered at an earlier position, its selected sentences will also appear at the earlier positions in the statement.
At last, we compose all the ordered sentence sequences of all research directions into the final statement.
%Because the statements tend to be long,  we reorder the sentences in  each cluster and pay extra attention to sentences from different  documents which do not have chronological ordering.
%We gather sentences  from the same document in a  set and independently sort sentences in every set by their chronological ordering.
%At last, we reorder different sets by greedily maximize the topical similarity of  their neighbor sentences.
%The case in Figure \ref{fig:reordering} illustrate this procedure.

It is noted that  we  adopt the summarization technique in sentence extraction and then order the extracted sentences according to their research directions which are determined by document clustering method. An alternative method is to first extract important sentences and  then cluster them. We do not adopt the alternative method because research directions should be determined by the content of papers, but not by sentences.
Another alternative method is to extract important sentences for each direction respectively. 
This method is not adopted because our preliminary experiments show that the clustering error may be propagated to the extraction step and degrade the whole performance.
Overall, our method is a practical solution to the RSG task at the limitation of our resources.
%At the same time, since our method is  the first attempt to this task, there is still a lot of improvement room.

\section{Experiments}

%\begin{table*}

%\centering
%\begin{tabular}{c|ccccccc}
%\hline
% &AP&AP(tf-idf)&AP(BERT)&K(c=2)&K(c=3)&K(c=4)&K(c=5)\\
% \hline
% BDI&\textbf{ 0.4099}&1.0905&1.1452&0.4325& 0.4927&0.4867&0.5892\\
 %CHI& 50.635&1.7487&3.5635&28.867& 50.983&\textbf{66.739}&36.609\\
%\hline
%\end{tabular}
%\caption{ Internal performance of AP algorithm with different similarity metrics comparing with K-Means (K) given different cluster  numbers.}
%\label{Clustering_result}
%\end{table*}

\subsection{Experiment Setup}
For the neural topic model, we select a vocabulary list of 2,000 most frequent words after removing the stop words.
We set the latent topic number to 100 and use Adam optimizer for training.
For sentence extraction, we fine-tune B\textsc{ertsum} on `bert-base-uncased'.
Using the crawler script for getting the CSPubSum corpus~\cite{collins2017supervised}, we get 8,953 scientific publications for training B\textsc{ertsum}\footnote{
B\textsc{ertsum} achieves outstanding performance as 0.47, 0.24 and 0.43 in ROUGE-1,2,L respectively on the CSPubSum test set of 149 publications after trained for approximately 30 minutes on a single NVIDIA 1080Ti}.
%after training for only about 30 minutes on a single NVIDIA 1080Ti
%To evaluate RSG performance, 
Our RSG dataset is divided into 31 statements with the corresponding papers for validation and the rest 31 for test. In this work, we only use the text from the \textit{Abstract} and \textit{Introduction} parts of the papers to generate the research statements. Table \ref{table:expdata} shows the data statistics.

\begin{table}
\centering
\small
\renewcommand\tabcolsep{0.85pt}
\begin{tabular}{c|ccc|ccc}
\hline
 & &Paper\#/statement& & &statement Len. \\
 & avg & max & min  & avg & max & min \\
\hline
Evaluation & 17.4 & 66 &  6 & 1912 & 6141 &  120\\
Test & 21.4 & 127 & 5  & 2023 & 5731 & 536 \\
 \hline
\end{tabular}
\caption{Statistics of RSG Data}
\label{table:expdata}
\vspace{-1.5em}
\end{table}

\subsection{Automatic Evaluation Metric}
For a fair comparison,  we set the length limit of the generated statement to 500 words as \citet{sun2019compare} suggested that it is unfair to use ROUGE-F to evaluate summaries with different lengths.
The metrics below are used to automatically compare our method with the baselines.

%\textbf{ROUGE}. We use the ROUGE metrics \cite{lin2004rouge}, including ROUGE-L and ROUGE-2, to measure the informativeness of the generated statements.When analysing the generated summaries,  we follow the evaluation way of long text generation \cite{liu2018generating} and only apply  ROUGE-F.
\textbf{ROUGE}. We follow the evaluation way of long text generation \cite{liu2018generating} and only apply  ROUGE-L \cite{lin2004rouge}  to measure the informativeness of the generated statements.

\textbf{B\textsc{ert}-S}. We use BERTS\textsc{ore}\footnote{https://github.com/Tiiiger/bert\_score} \cite{zhang2019bertscore} to evaluate semantic similarity of the generated statements with their references.
XLNet \cite{yang2019xlnet} based BERTS\textsc{ore}\footnote{The default BERT based BERTS\textsc{ore} has the restriction of 512 words for the input length. } is applied here.
%Because more than half of the reference statements are longer than 512 words restriction of default BERT based BERTS\textsc{ore}, we apply XLNet \cite{yang2019xlnet} based BERTS\textsc{ore} to model arbitrary input length.

\textbf{Entity Recall (ER)}. 
We design a new metric ER to measure the overlap percentage of scientific terms that appear both in the reference and generated statements, as a good generated statement should share more common  scientific terms with the reference.
%This metric is used to measure how many scientific terms in the reference statements are also mentioned in our generated statements. 
%We use ER to measure how many scientific terms in reference statements are also mentioned in our generated statements. 
%ER is defined as the percentage of scientific terms in reference statements that also appear in  generated statements.
To extract scientific terms, we use the NER model Scibert\footnote{https://github.com/allenai/scibert} \cite{beltagy2019scibert}, which is pretrained on scientific corpus and tuned on the NER dataset SciERC \cite{luan2018multi}.
%We use the state-of-art NER model Scibert \cite{beltagy2019scibert}, which is pretrained on scientific corpus and tuned on computer science domain NER dataset SciERC \cite{luan-etal-2017-multi},  to extract scientific terms.

We use the AP method to determine the research directions. To evaluate its performance, we adopt the Davies-Bouldin Index (DBI): a commonly used clustering evaluation metric.
A lower DBI value means that the model can better separate the clusters.
That is, the papers being clustered into the same research direction are more similar and different directions are well separated.

\subsection{Human Evaluation Metric}
Since it is difficult to automatically evaluate text organization and language quality, we also manually measure the generated statements.
Here we choose two metrics  of content coverage and text coherence.
Content coverage (CC)  measures whether the research statement describes all the research directions discussed in the reference statement by providing concise and informative sentences.
Text coherence (TC) mainly evaluates  whether the research statement is well-organized in different directions and whether the text describing each research direction is coherent.
Three volunteers with academic background in computer science  score the statements from 1$\sim$10 for each metric. The higher the score, the better the statement is.

%\subsection{Model Comparison}
\subsection{Method Analysis}

\begin{table}
\centering
%\small
\begin{tabular}{l|cccc}
\hline
Model&ROUGE-L&B\textsc{ert}-S&ER\\
\hline
\textbf{ORACLE}&\textbf{52.67}&\textbf{72.17}&\textbf{15.4}\\
Random&24.12&57.98&2.26\\
Multi-Lead&25.20&60.01&3.92\\
TextRank&25.39&60.12&3.47\\
LexRank&25.72&59.31&3.34\\
SUMO &25.32&59.61&4.15\\
%SUMO$_{+MMR}$ &25.18&59.34&4.23\\
%PG(Random)&26.13&60.64&3.16\\
%PG(TextRank)&26.24&60.73&4.91\\
%SUMO&25.3&59.6&4.15\\
%SUMO+MMR&25.2&59.3&4.23\\
\hline
%B\textsc{ertsum}+MMR&\textbf{27.7}&\textbf{60.8}&\textbf{4.96}\\
Ours$_{-MMR}$ &27.67&60.74&4.81\\
Ours &\textbf{27.74}&\textbf{60.77}&\textbf{4.96}\\
\hline
\end{tabular}
\caption{Automatic Evaluation of Methods}
\label{table:exp1}
\end{table}

%To evaluate the RSG performance, we evaluate each module of our RSG method by comparing with some representative baselines.% or designing different settings.
To evaluate our  RSG method, we test  each module 
by comparing with some representative baselines.

\subsubsection*{Summarizer Module}
%Our automatic evaluation results on RSG are shown in Table \ref{table:exp}.
%The results of our model with various settings and other carefully selected representative baselines are listed in Table \ref{table:exp}.
%To evaluate the RSG performance, we compare our method with some representative baselines.
We first evaluate the summarizing module. % by comparing with some representative baselines.
Ours is our adopted summarizer which is trained on CSPubSum.
textit{ORACLE} greedily selects the the sentences that are most similar to the ground-truth  statements, and can be seen as the upper bound of an extractive method for RSG. 
%\textit{ORACLE} greedily selects the sentences that are most similar to the ground-truth  statements， and can be seen as the upper bound of an extractive method for RSG. 
The \textbf{Random} method randomly selects some sentences from the  papers as the statement.
\textit{Multi-Lead} picks out the lead sentences from each paper's abstract and combines them into a research statement.
\textit{TextRank} \cite{mihalcea2004textrank} and  \textit{LexRank} \cite{erkan2004lexrank} are two unsupervised  extractive summarization methods which are based on sentence similarity and graph-based ranking algorithms.
%These two methods work  on the segmented sentences from the papers only and do not exploit any external resources.
%We evaluate the  performances and limitations of only applying unsupervised extractive methods without external corpus.
%\textit{TextRank} builds on an undirected graph while \textit{Pacsum} improves \textit{TextRank}by building directed graph and uses BERT-based similarity computation.
%\textit{Pacsum}(tf-idf) adopts tf-idf features to compute sentence similarity on an undirected graph and \textit{Pacsum}(BERT) further uses fine-tuned BERT representations to compute similarity. In this work, we use  CSPubSum for BERT to learn the similarity metrics.
%We experiment \textit{Pacsum}(tf-idf) with both tf-idf features and BERT-based similarity trained on CSPubSum according to its original paper.
%\textit{TextRank} is a well-known graph-based method and \textit{Pacsum} is a recently proposed centroid based method that performs better than other unsupervised extractive methods.
%We experiment \textit{Pacsum} with both tf-idf features and BERT-based similarity trained on CSPubSum according to its original paper.
%We experiment \textit{Pacsum} with both tf-idf features and BERT-based similarity trained on CSPubSum according to its original paper.
%It does not show much of improvements using BERT in our experiment.

%\textit{SUMO} is a relatively weaker summarizer compared to B\textsc{ertsum}.
%We train \textit{SUMO} on  CSPubSum, the same with B\textsc{ertsum}.
%From this baseline, we demonstrate how the performance of the summarizer module affect total performance.
\textit{SUMO} is  a recent supervised extractive summarization model , which  induces a sentence-level tree structure for one document and predicts the root node as the summary based on the Transformer architecture~\cite{liu2019single}. 
We also train \textit{SUMO} on CSPubSum.

Table \ref{table:exp1} shows the  results of all methods with respect to ROUGE-L, BERT-S and ER.
%We can see that B\textsc{ertsum} outperforms all the baselines. 
%Concatenation of Random sampled summaries achieve a better performance in Rouge-L than other models beside our model but a lower ER which implies the negligence of important scientific terms.
As we expect, \textit{Random} performs the worst among all the methods. Especially, its low ER value implies its negligence of extracting important scientific terms.
Unlike \textit{Lead-3} which extracts the first 3 sentences and achieves a good performance on news summarization, the performance of \textit{Multi-Lead} on RSG is mediocre, though much better than \textit{Random}.
%lower than most of the other baselines. 
%Though not comparable with other methods, \textit{Multi-Lead} is still much better than \textit{Random}.
This implies that %sentences from 
abstracts contain more useful information than the other parts of papers, but the first sentences from abstracts usually talk about some general background and may cause redundancy when summarizing papers with similar topics.

We also observe that the supervised model \textit{SUMO} performs almost on a par with the two unsupervised methods \textit{TextRank} and \textit{LexRank} with regard to ROUGE-L and BERT-S, but better on the ER metric. We infer that \textit{SUMO} may benefit from supervision of  CSPubSum which can help to capture more scientific terms, while \textit{TextRank} and \textit{LexRank}  score sentences based on sentence similarity which do not distinguish between scientific and non-scientific terms.
Our method unsurprisingly performs better than 
%the transformer-based 
\textit{SUMO}, showing the power of BERT in text representation.
%The abstractive method PG(TextRank)  shows its potentials with a large training data and achieves a similar performance with B\textsc{ertsum} concerning B\textsc{ert-s} and ER. As most sentences in the statement are slightly rewritten from the sentences in the original papers, the generated sentences of PG may degrade the ROUGE value which measures the word sequence overlap.
From the bottom block of Table \ref{table:exp1}, we can also find that  MMR can improve the overall performance  by penalizing redundancy.

%We also observe that our method performs better than the unsupervised summarization methods \textit{TextRank} and \textit{LexRank}, meaning that the corpus of CSPubSum is capable of giving some supervision in sentence selection because CSPubSum is composed of scientific publications and can help to capture some efficient features for scientific summarization.
%The selection of training corpus is important, which is also supported by the poor performance of \textit{PG} supervised by the news corpus CNN/DM. 
%Our method also performs better than \textit{SUMO} which is based on the transformer framework, showing again the power of BERT in the text representation.
%%%说一下为什么SUMO比无指导模型差almost on a par with the overall performance***
%%%说一下MMR的作用
%%%和oracle相比
%The bottom block of Table \ref{table:exp1} shows the RSG performance without MMR.
%From the table, we can observe that MMR improves the ER metric by penalizing redundancy.

%Using TextRank to further select generated summaries significantly improves ER and slightly improves ROUGE-L and B\textsc{ert}-S and achieve a close performance to our model which indicates  the potentials of generation methods but it require multiple times of examples and time to train.
%sWe left abstractive method for our task as future work. 

Through model comparison, we can see that our method is an acceptable and practical solution to the RSG task, though its overall performance is still far from \textit{ORACLE} and can be further improved. 
%by truncating each input paper to 400 words and concatenate the output summaries until reaching the maximum length.
%The result of PG is worse than extractive methods by a large margin because of its deficiency in encoding long paragraphs.

%\subsection{Module Analysis}
\subsubsection*{Clustering Module}

\begin{table}
\centering
%\small
\begin{tabular}{l|c}
\hline
&DBI\\
\hline
AP&\textbf{ 0.4099}\\
AP(tf-idf)&1.0905\\
AP(BERT)&1.1452\\
K-Means(K=2)&0.4325\\
K-Means(K=3)&0.4927\\
K-Means(K=4)&0.4867\\
K-Means(K=5)&0.5892\\
\hline
\end{tabular}
\caption{Comparison of clustering methods with the DBI metric}
%Internal performance of AP algorithm with different similarity metrics comparing with K-Means  given different cluster  number $c$.}
\label{Clustering_result}
\end{table}

\begin{figure}
\centering
\includegraphics[scale=0.15]{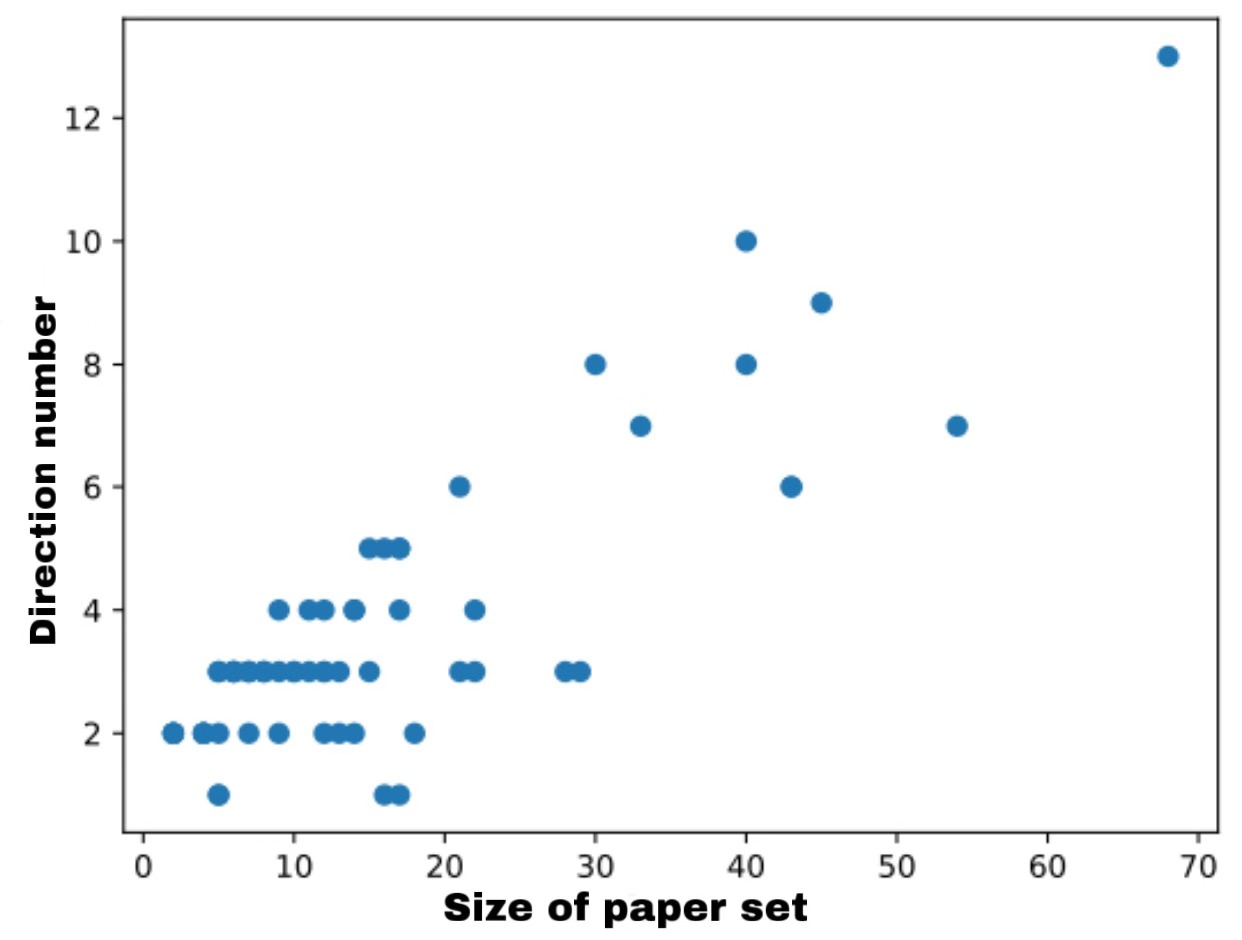}
\caption{Statistics of research directions by AP clustering method}
%Each point in the scatter plot represents an author.  }
\label{fig:clustering}
\end{figure}
We also perform extra evaluations on our clustering module.
As the AP clustering method automatically determine the number of research directions for each researcher, we show the statistics of research directions as in Figure \ref{fig:clustering}. We can see that the researcher with more papers tends to be involved in more research directions and most  researchers have less than 6 directions.
%As we use the nonparametric AP clustering method, the first essential thing is the rationality of the number of clusters determined by the clustering method itself.
%Figure \ref{fig:clustering} shows the statistics after clustering, the author with more papers tends to have more directions to discuss. For example, an author of more than 40 papers has 6 directions and an author of 20 papers may have 3 to 4 directions.
%It's also noticed that the number of directions is in an acceptable range.
%Most of the researchers have less than 6 directions after clustering.
%Since the effects of the clustering method can not be reflected by the ROUGE metrics, 
We use the DBI metric to compare the performance of some typical clustering methods to verify the effectiveness of choosing AP clustering with topical similarity (AP).
AP(tf-idf) and AP(BERT) uses tf-idf and BERT (without fine-tuning)  representations for similarity computation respectively. $K$-Means methods with different $K$ values are also used for comparison. From Table \ref{Clustering_result}, we can find that $K$-Means tends to keep the cluster number between 2 and 4, which is consistent with the clustering results by AP.  AP with topical similarity achieves the best performance, indicating that it is effective to automatically determine the cluster number and topical representation is suitable to similarity measurement in our task.
%Our clustering internal performance are listed in Table \ref{Clustering_result} 
%where AP(tf-idf) and AP(BERT) uses product between tf-idf and BERT  (without fine-tuning)  vector for similarity metric.
%The rest of the models use topical similarity.
%The results indicate that topical similarity is a suitable similarity measurement for document clustering in our task.
%AP algorithm with topical similarity achieves the best performance. %, and a relatively close performance in CHI metric to K-Means model with 3 clusters which is considered an acceptable cluster number.
%Overall, our clustering module achieves competitive performance and more flexible without a pre-defined cluster number.
%\subsection{Human Evaluation}
\begin{table}
\centering
%\small
\begin{tabular}{c|cc}
\hline
Model&CC&TC\\
\hline
Abs-Comb&\textbf{6.17}&4.83\\
\hline
%B\textsc{ertsum}$_{CSP}$+MMR&6.08&\textbf{5.92}\\
Ours&6.08&\textbf{5.92}\\
%\hline
 \textbf{w/o }reordering&6.08&4.92\\
%\hline
 \textbf{+w/o }clustering&5.92&4.50\\
 \hline
\end{tabular}
\caption{Human evaluation on sentence clustering and reordering}
\label{table:exp}
\end{table}

Further, we conduct human evaluation to measure the effects of the clustering and reordering modules. 
The results are shown in Table \ref{table:exp},
A strong baseline is named \textbf{Abs-Comb} which combines all the corresponding abstracts into a research statement.
%In the human evaluation, we compare our model with the strong baseline \textbf{Abs-Comb} which combines all the papers' abstract as research statements.
For the fairness of comparison, we use our model  to extract sentences until we have the same number of sentences as \textbf{Abs-Comb}.
%we use our best-performed model SDT+MMR with AP clustering to extract until the same number of sentences as \textbf{Abs-Comb}.
\textbf{w/o reordering} only removes the sentence reordering process which orders sentences from different papers and \textbf{+w/o clustering} further removes the clustering module.
 %Our results are shown in Table \ref{table:exp}, \textbf{w/o reordering} only removes the sentence reordering process which orders sentences from different papers and \textbf{+w/o clustering} further removes the clustering module.
Concerning the metric of content coverage (CC), all the models perform nearly the same. \textbf{Abs-Comb} can cover a little more content than our method while sacrificing text coherence(TC).
%\textbf{Abs-Comb}  covers all the topics of the statement but it introduces redundancy from the same topic papers.
For example, similar sentences may appear several times in the results of \textbf{Abs-Comb}.
%if an author published several papers in a similar area.
We can also see that text coherence has dropped considerably from 5.92 to 4.92 without reordering. Without clustering, the performance continues to decline from 6.08 to 5.92 in CC and from 4.92 to 4.50 in TC, meaning that
clustering is helpful to organizing the sentences. It is interesting that \textbf{+w/o clustering} does not reduce the selected sentences but makes a worse impression in content coverage compared to the results with clustering.
%Though in automatic evaluation models without clustering achieve a higher ROUGE score, in human evaluation when the paragraph becomes much longer, one may consider better-organized documents to be more informative.

%For topical coherence,  \textbf{Abs-Comb} and \textbf{w/o clustering} only concatenate selected sentences from different papers providing non-topic at all.
%All the raters agree that the clustering step and reordering step improve the statement's readability and coherence, which is reflected from the improvements in the \textbf{TC} metric.
%Specifically, the sentence reordering procedure after clustering significantly improves the coherence according to the results.
%Overall, our full model provides much better statements according to the average score  \textbf{Ave}.
\begin{figure}

\includegraphics[scale=0.25]{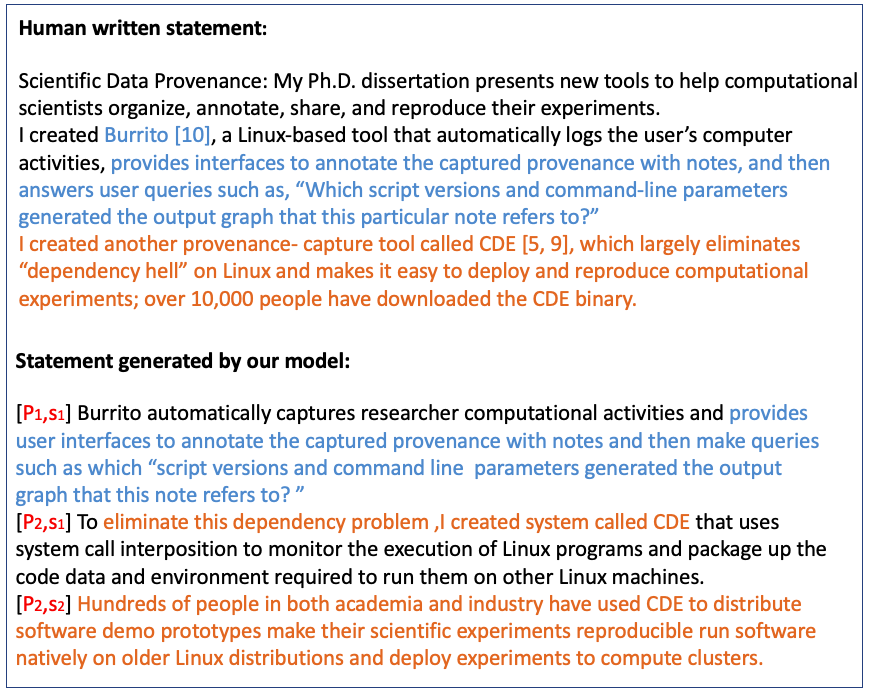}
\caption{Case study of research statement generation }
\label{fig:case}
\end{figure}
\subsection{Case Study}
We illustrate two intuitive examples to show the performance of our RSG method.
Figure \ref{fig:case} displays a part of the system generated statement as well as the corresponding human  statement. 
The human statement contains three sentences summarized from three papers with two contributions which are highlighted by different colors.
We can observe that our model has selected three sentences  which can cover the contributions included in the human statement.
That is, our extractive method can well pick out the researchers' contributions from the papers.
Meanwhile, our method is limited to losing some content generalized by the researchers (e.g., the first sentence in the human statement). 

%We illustrate a part of our generated statement comparing to the corresponding part of the human written statement in Figure \ref{fig:case} and showing the effect of sentence reordering in Figure \ref{fig:reordering}.  
%The human written statement in Figure \ref{fig:case} contains three sentences containing two contributions that correspond to three papers.
%Our model selects three sentences from these three papers which cover all the information in the researcher's contributions as we underlined.
%Overall, we completely summarize the researcher's contribution sentences due to the advantage of the extractive method.
%Meanwhile, because of the limitation of the extractive method, we lose the abstractive first sentence in the human written statement.
\begin{figure}
\centering
\includegraphics[scale=0.28]{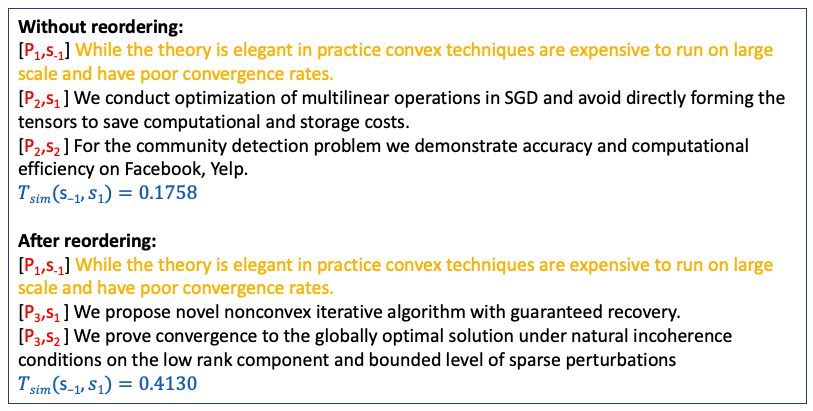}
%\caption{Illustrating the reordering module.}
\caption{Case study of text reordering.}
\label{fig:reordering}
\end{figure}

Figure \ref{fig:reordering} shows an example of two sentence sequences before and after text reordering which involves the research direction of \textit{PCA}. 
The  first sentence $s_{-1}$ , which is the last sentence of the paper $P_1$, describes the disadvantages of  \textit{convex techniques}. 
Without reordering, this sentence is randomly succeeded by the first sentence of paper $P_2$ which discusses application in \textit{convex techniques} given the topical similarity of the two sentences is 0.17. 
After reordering, [$P_1$,$s_{-1}$] is right before the first sentence of paper $P_3$ which talks about a method that the author proposed to solve the disadvantages of  \textit{convex techniques}.
This example shows that text reordering based on topical similarity can well improve the text coherence.
%which involves the effect of sentence reordering. In t

%Figure \ref{fig:reordering} is sampled from the direction about \textit{PCA} from a generated statement.
%The first sentence $s_{-1}$ , also the last sentence of $P_1$, describes the disadvantages of  \textit{convex techniques}.
%Without reordering, it randomly combines with the first sentence of paper $P_2$ which discussing application in \textit{convex techniques} given the topical similarity of about 0.17. 
%After reordering, it matches the first sentence of paper $P_3$, by a much higher topical similarity, which talking about a method the author proposed to solve the disadvantages of  \textit{convex techniques}.
%The comparison shows obvious improvement in coherence.  

\section{Related Work}

Research statement generation is closely related to MDS techniques and should also care for the characteristics of scientific publications. 
%Here we briefly introduce some work of MDS and scientific summarization.

\subsection{Multi-Document Summarization}
MDS is pioneered by the work of \cite{mckeown1999generating} and other early notable work includes \cite{mckeown1999towards,radev2004centroid}.
%Most MDS systems apply extractive methods~\cite{cao2015ranking,peyrard2017supervised} which produce a summary by directly selecting a number of important sentences from multiple input documents.
For a long time, the mainstream MDS methods have been extraction based ones ~\cite{wan2007manifold,cao2015ranking,peyrard2017supervised} which produce a summary by directly selecting a number of important sentences from multiple input documents. 
Usually, these models are composed of the two steps of sentence scoring and sentence ordering which are based on various kinds of machine learning techniques.
Redundancy is one of the major problems in MDS and a well-known method for this problem is Maximal Marginal Relevance \cite{carbonell1998use} and also recent DPP based methods \cite{kulesza2011learning,cho2019improving}.
%Recent works DPP based methods \cite{kulesza2011learning} consider both sentence quality and similarity in sentence extraction provide a promising result. 
With the development of sequence-to-sequence neural networks, some studies have attempted abstractive methods on MDS \cite{zhang2018towards,lebanoff2018adapting} and a large news    dataset Multi-News \cite{fabbri2019multi} has been proposed.
New MDS tasks like generating Wikipedia pages ~\cite{liu2018generating,liu2019hierarchical,li2020leveraging} and unsupervised abstractive MDS \cite{chu2018meansum,bravzinskas2019unsupervised} also attract much attention.

\subsection{Scientific Summarization}
There are two types of summarization tasks for scientific publications:  article abstract generation and citation-based summarization \cite{cohan2017scientific}.
Article abstract generation aims to generate a  summary for the article which may be better than the original abstract ~\cite{elkiss2008blind}.
%Recently, \cite{collins2017supervised} have built two large datasets CSPubSum and CSPubSumExt  for scientific publication summarization.
Recently, some large dataset like CSPubSum \cite{collins2017supervised}, PubMed and arxiv \cite{cohan2018discourse} for scientific publication summarization have build scientific publication summarization.
The citation-based summarization method aims to summarize the content of a set of citations to a referenced article \cite{qazvinian2008scientific,qazvinian2013generating}. How to make use of citation to supplement the statement will be our future consideration. 
%For future work, we plan to fetch citation sentences as additional input and incorporate citation summarization module to our RSG model.

%The citation-based summarization method mainly generates a summary by extracting a set of citations to a referenced article \cite{qazvinian2008scientific,qazvinian2013generating}.

\section{Conclusion and Future Work}
In this paper, we propose the research statement generation (RSG) task which aims to summarize one's research achievements and help prepare a formal research statement.
For the RSG task, we propose a feasible method which uses topic modeling and AP clustering method to determine research directions and  BERT-based summarization method to extract salient sentences.
%and  experiment with some representative recent methods as baselines. 
%We train a neural topic model for getting topical similarity for documents and sentences, based on which we cluster the published papers into several research directions using the affinity propagation clustering algorithm.
%Then we  summarize every direction by the ertSum and learn to select the sentences which most reflect the researcher's contributions.
%Finally, we reorder the selected sentences from different documents for better text coherence.
Our method is a first attempt on the RSG task and expects to inspire more efficient methods for reducing the efforts of writing a research statement.

In future work, we will conduct further research in two aspects. 
First, we will explore abstractive methods to further improve RSG performance.
Second, we will introduce more evidence about a researcher's contributions such as personal information and citations.

\newpage

\section*{Acknowledgements}

\bibliography{anthology,custom}
\bibliographystyle{acl_natbib}

\appendix

\section{Example Appendix}
\label{sec:appendix}

This is an appendix.

\end{document}